\documentclass{hep99}
\usepackage{epsfig}

\begin{document}

\title{The Partonic Structure of the Quasi-Real Photon}

\author{J.H. Vossebeld (On behalf of ZEUS and H1)}

\address{%On behalf of the ZEUS and H1 collaborations\\
NIKHEF, P.O. Box 41882, 1009 DB Amsterdam, The Netherlands\\
E-mail: {\tt vossebeld@nikhef.nl}}

\abstract{New measurements of dijet photoproduction at HERA provide information on the gluon density in the photon and on its quark densities at high $x_\gamma$ and high factorisation scales.} 

\maketitle

\section{Introduction\footnote{The results discussed here have also been described in the papers 157ad and 540, submitted to this conference.}}
In dijet photoproduction two event classes are distinguished: 
direct events, in which the photon couples directly 
to a parton in the proton, and resolved events, in which the 
photon acts as a source of partons, one of which scatters off 
a parton in the proton. Resolved processes are sensitive to the 
partonic structure of the photon. 
Experimentally direct and resolved events can be 
separated using $x_\gamma^{obs}\,(=x_{\gamma,jets}$), the fractional 
momentum of the photon participating in the hard scatter, defined as:
$x_\gamma^{obs}={(E_{T\,1}^{jet}e^{-\eta_1^{jet}}+E_{T\,2}^{jet}e^{-\eta_2^{jet}})}/{2yE_e}$, 
%\[x_\gamma^{obs}=\frac{E_{T\,1}^{jet}e^{-\eta_1^{jet}}+E_{T\,2}^{jet}e^{-\eta_2^{jet}}}{2yE_e}\,,\]
where $y$ is the fractional momentum of the incoming positron transferred 
to the photon. 
%Dijet photoproduction measurements at HERA can provide information on the gluon density in the photon and on the quark densities at high $x_\gamma$ and high scales. 

\section{Measurement of the gluon density in the photon}
H1 has measured the dijet photoproduction cross section as a function of 
$x_{\gamma,jets}$, using $7.2\,\rm{pb}^{-1}$ of data collected in 1996.
From this measurement the gluon density in the photon is extracted, using 
the {\it single effective subprocess} approximation~\cite{ses}, in which 
the cross section factorises in an effective scattering 
process and effective parton densities in the proton and the 
photon, defined by:
$f_{eff}=\sum_i^{n_f}{(q_i+\bar{q}_i)}+\frac{9}{4}g$, 
%\[f_{eff}=\sum{(q+\bar{q})}+g\,,\]
where $n_f$ is the number of active quark flavours.
 
Jets are reconstructed using a cone algorithm with cone radius $R=0.7$. 
The sensitivity to multiple interaction effects is reduced by subtracting 
pedestals from the transverse momenta of jets. A kinematic
region is selected satisfying: $-0.5<\eta^{jets}<2.5$, $P_T^{jets}>6$~GeV, 
$\vert\eta_1^{jet}-\eta_2^{jet}\vert<1$, $0.5<y<0.7$ and 
$Q^2<0.01$~GeV$^2$.

\begin{figure}[h]
\begin{center}
\begin{picture}(100,67)(0,0)
\put(0,-19.5){\includegraphics[width=1.45in]{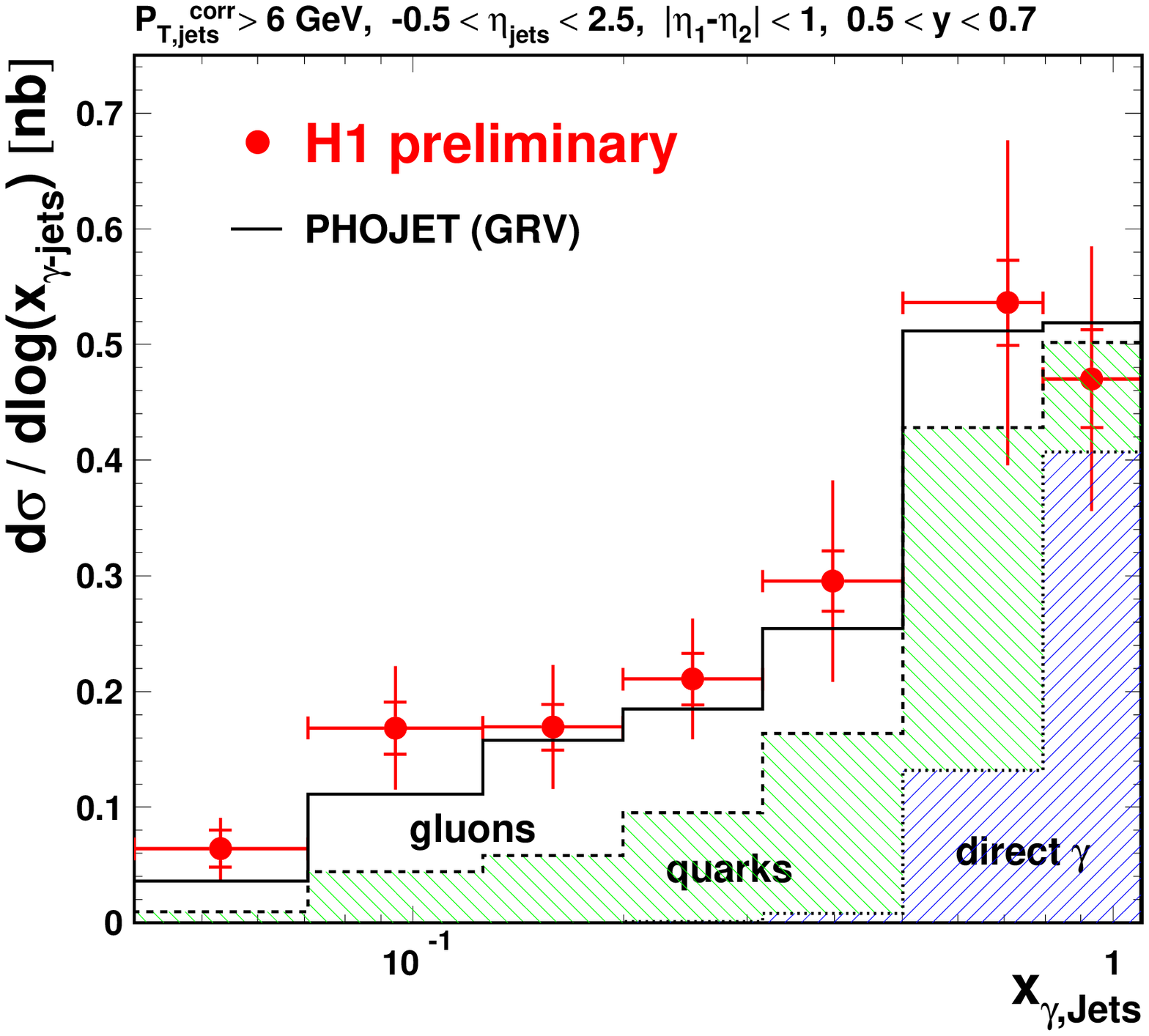}}
\end{picture}
\end{center}
\caption{{\footnotesize Dijet cross section as a function of $x_{\gamma,jets}$ compared to the PHOJET Monte Carlo, using the GRV parametrisation of the photon structure.}}
\vspace{-.6cm}
\label{fig:dsigdxg}
\end{figure}

In figure \ref{fig:dsigdxg} the cross section is compared to the  
PHOJET~\cite{phojet} Monte Carlo, using the GRV~\cite{grv} 
parametrisation for the photon structure. 
Direct events and resolved events due to quarks and gluons
in the photon are shown separately. 
At low $x_{\gamma,jets}$ a clear sensitivity of the cross section to the 
gluon content of the photon can be observed.

\begin{figure}[b]
\begin{center}
\begin{picture}(228,60)(0,0)
\put(0,-17){\includegraphics[width=1.53in]{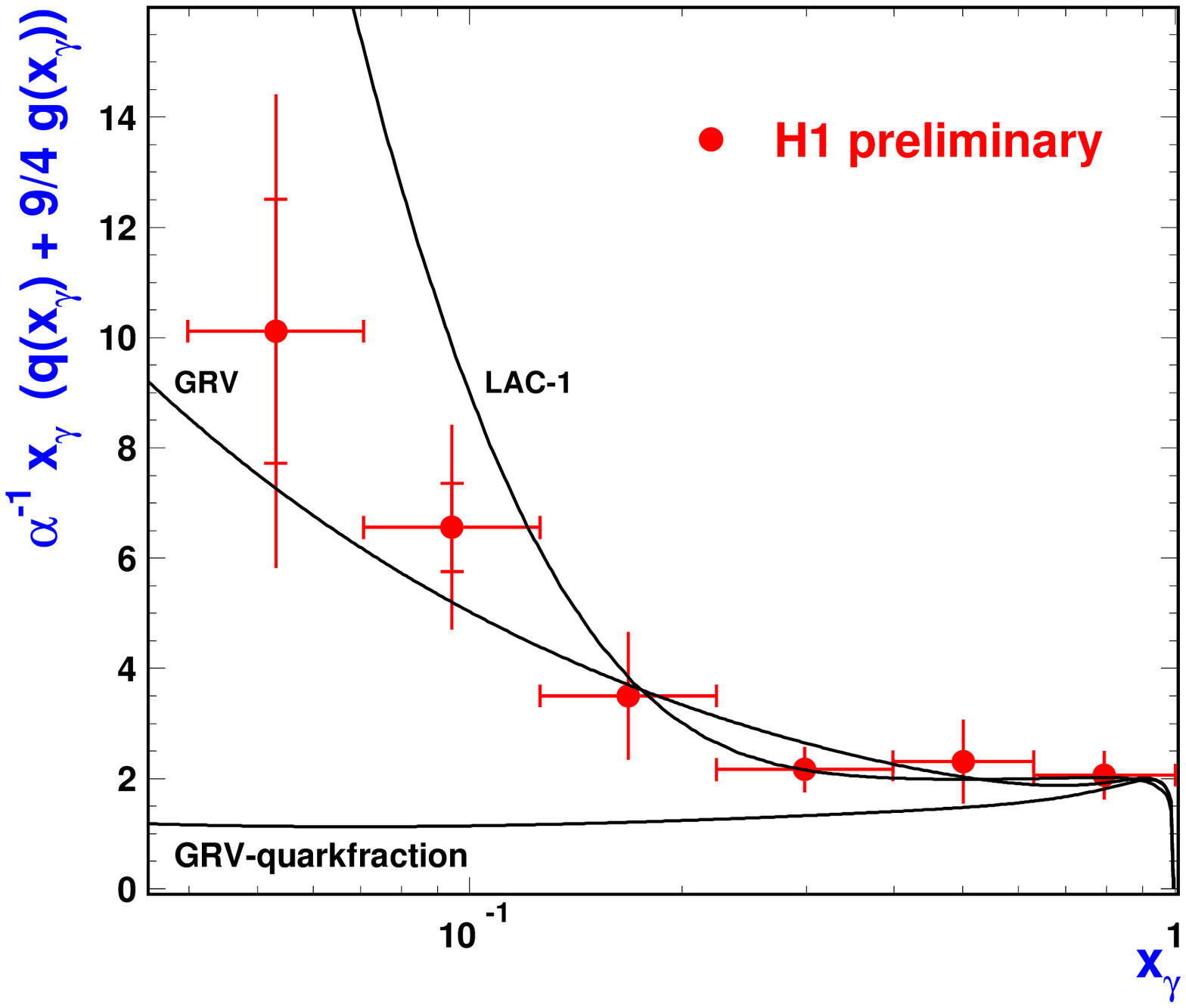}}
\put(112,-17){\includegraphics[width=1.53in]{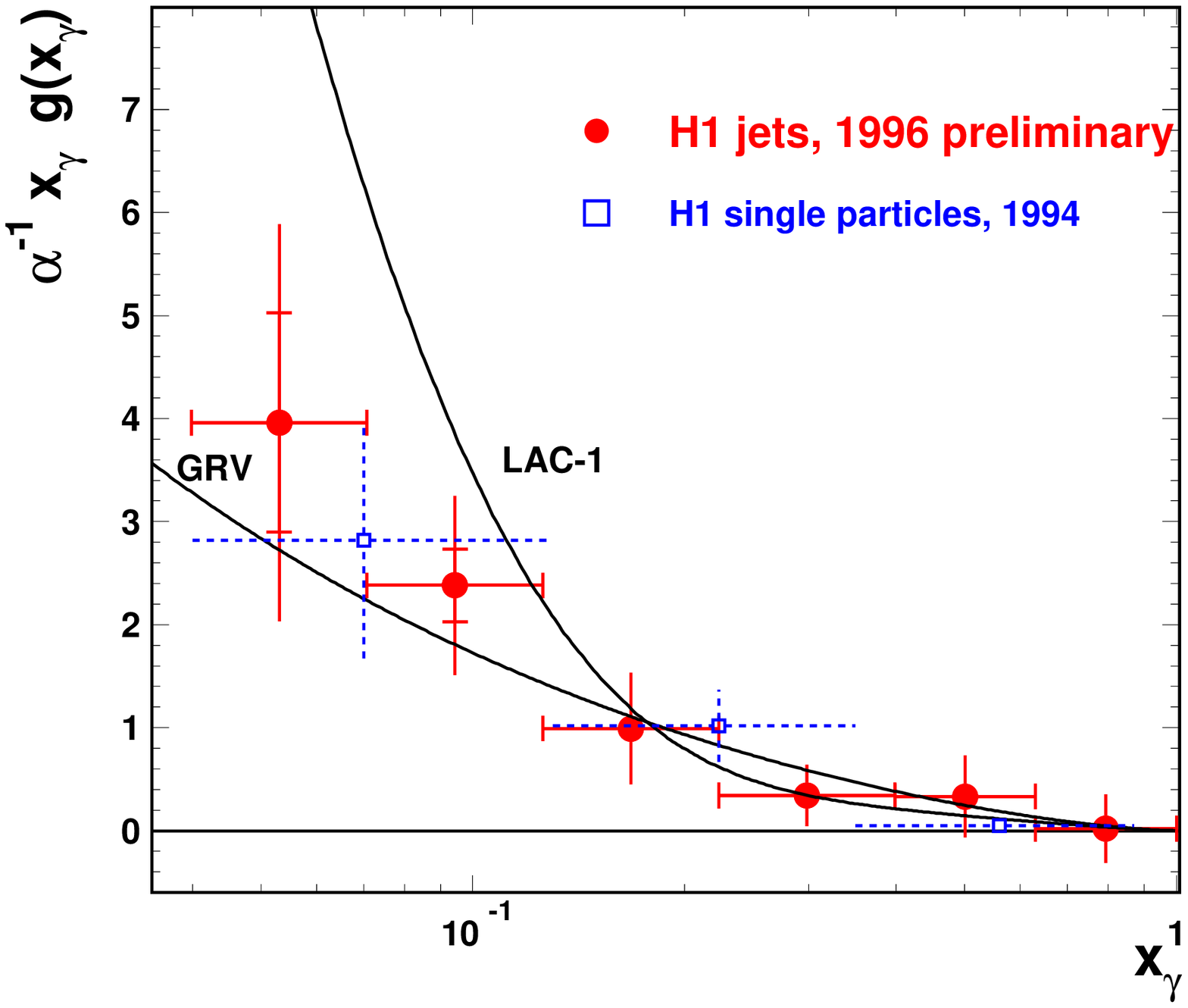}}
\scriptsize{\put(16,65){(a)}
\put(130,65){(b)}}
\end{picture}
\end{center}
\caption{{\footnotesize  The effective parton density (a) and the gluon density (b) in the photon compared to GRV and LAC1. The open squares in figure (b) were determined in a measurement~\cite{h1_chargedpart} of the charged particle cross section in photoproduction.}}
\label{fig:gluon1}
\end{figure}

After unfolding to the parton level, the effective parton density in 
the photon is determined from the ratio between the cross sections  
in data and Monte Carlo. 
Quark densities, as given by GRV, which agree with existing 
$F_2^\gamma$ data~\cite{f2gamma}, are subtracted to obtain the 
gluon density. 
The effective parton density and the gluon density, 
shown in the figures 
\ref{fig:gluon1}a and \ref{fig:gluon1}b, are compared to the 
GRV~\cite{grv} and LAC1~\cite{lac} densities. The gluon density  
rises at low $x_\gamma$ and favours the low GRV gluon density. 
Figure \ref{fig:gluon1}b also shows results obtained 
from a measurement of the charged particle cross section in 
photoproduction~\cite{h1_chargedpart}.

\section{Measurement of dijet photoproduction at high transverse energies}

ZEUS has measured the dijet cross section as a function 
of the highest transverse jet energy  
and the jet pseudorapidities, using $38\,\rm{pb}^{-1}$ of data collected 
in 1996 and 1997. 
A kinematic region is selected where:
$-1<\eta_{1,2}^{jet}<2$, $E_{T\,1}^{jet}>14$~GeV,  
$E_{T\,2}^{jet}>11$~GeV, $0.20<y<0.85$ and $Q^2<1$~GeV$^2$.
Jets are reconstructed using a $k_T$ clustering algorithm~\cite{kt1} 
in the inclusive mode~\cite{kt2}.

\begin{figure}[b]
\begin{center}
\begin{picture}(228,73)(0,18)
\put(0,0){\includegraphics[width=1.5in]{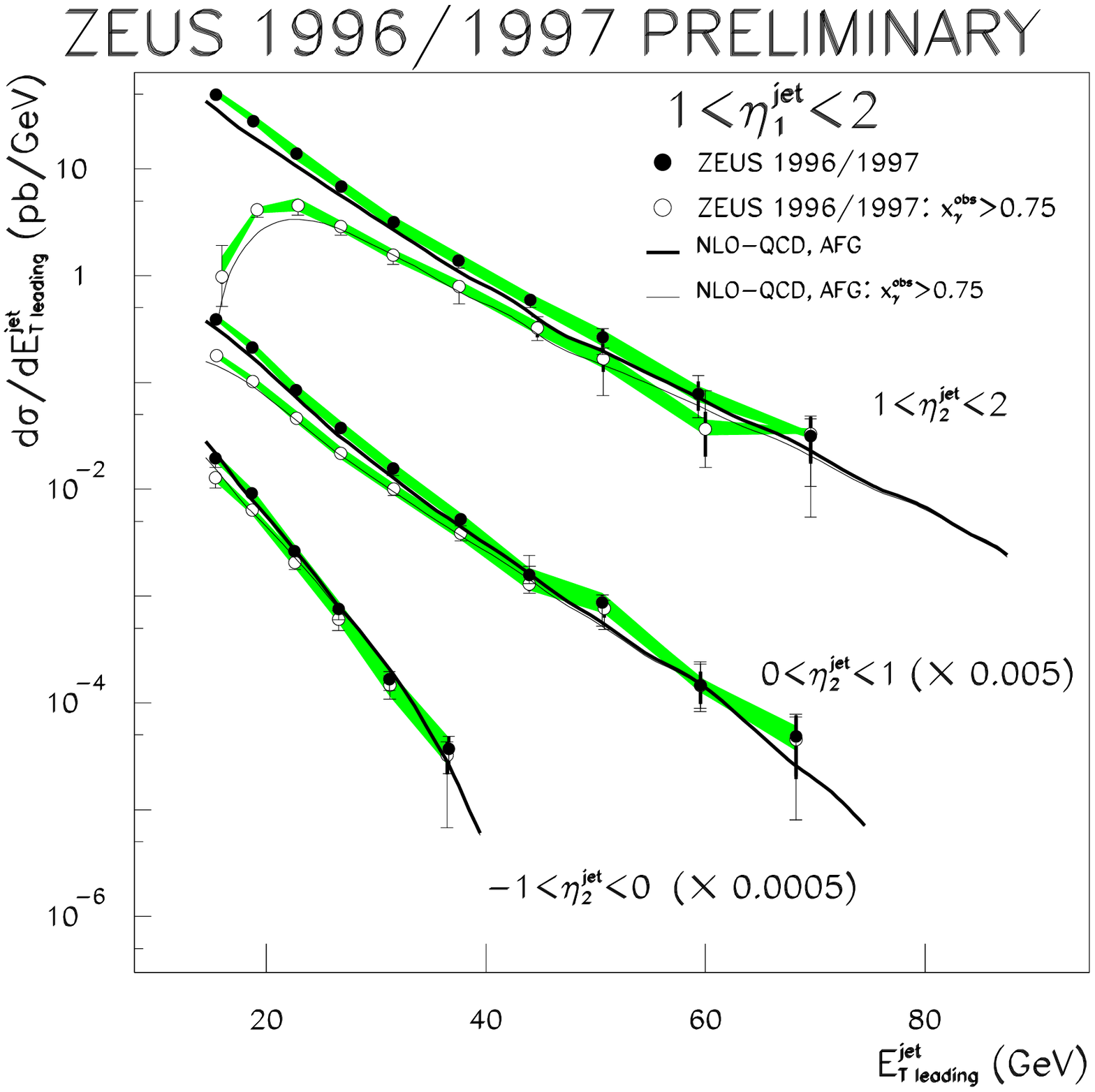}}
\put(116,0){\includegraphics[width=1.5in]{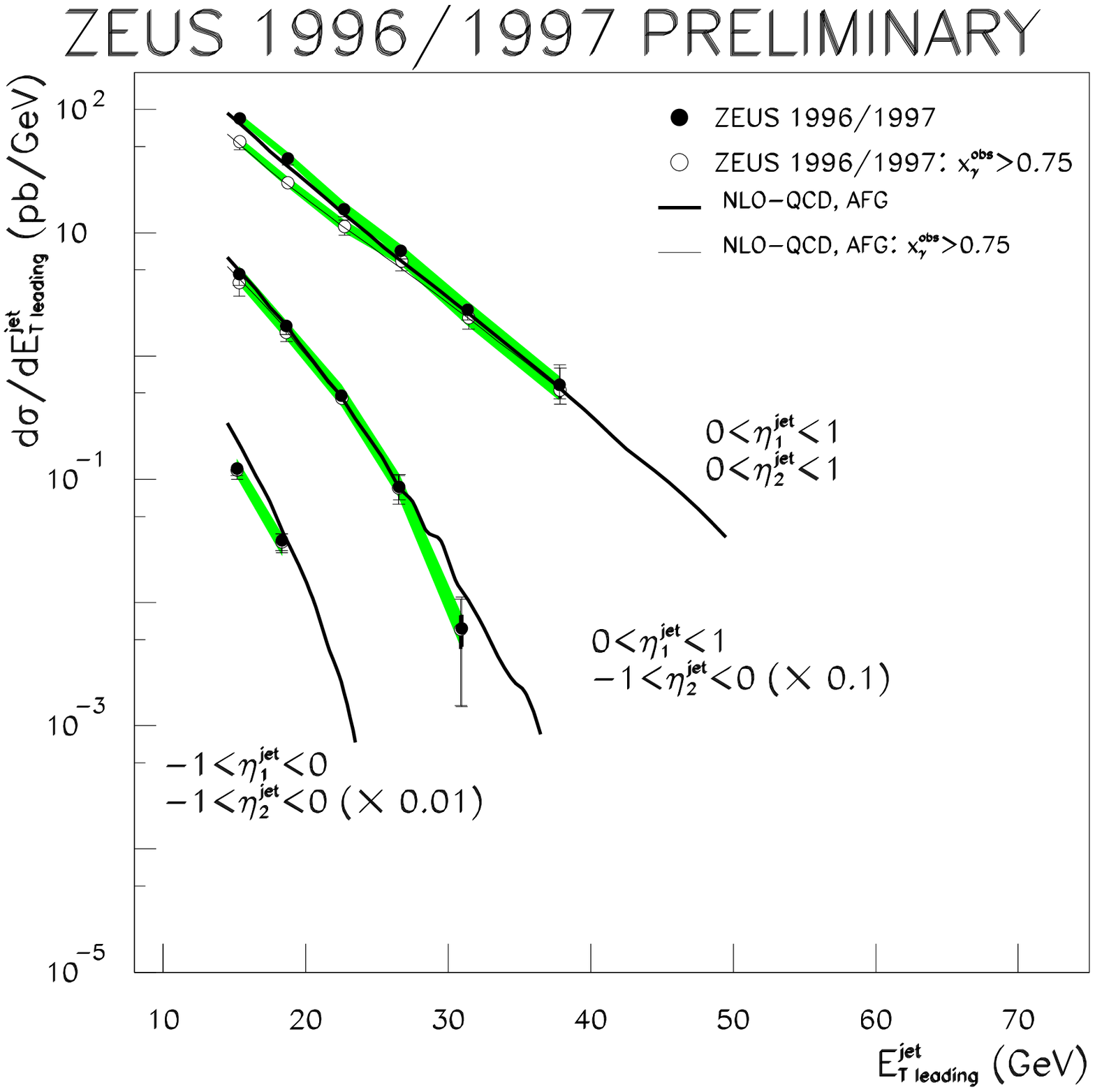}}
\end{picture}
\end{center}
\caption{\footnotesize Dijet cross section as a function of the highest transverse jet energy, in various regions of the jet pseudorapidities, 
for the full $x_\gamma^{obs}$ range and for $x_\gamma^{obs}> 0.75$, compared to NLO QCD. Thick error bars represent the statistical uncertainty and thin error bars the systematic and statistical uncertainties added in quadrature. The shaded band represents the energy scale uncertainty.}
\label{fig:dsigdet}
\vspace{-.6cm}
\end{figure}

Figure \ref{fig:dsigdet} shows the cross section as a function of the 
highest transverse jet energy in various regions of the jet pseudorapidities, 
compared to NLO QCD calculations~\cite{nlofr} using the AFG-HO~\cite{afg} 
parametrisation of the photon structure. The cross section was measured 
for the full data sample and for a direct-enriched sample with 
$x_\gamma^{obs}>0.75$. In general, data and theory agree, 
with the exception of the forward region, to be discussed below, 
where the data lie above the predictions.

\begin{figure}[h]
\begin{center}
\begin{picture}(185,175)(0,15)
\put(0,160){\includegraphics[width=2.56in]{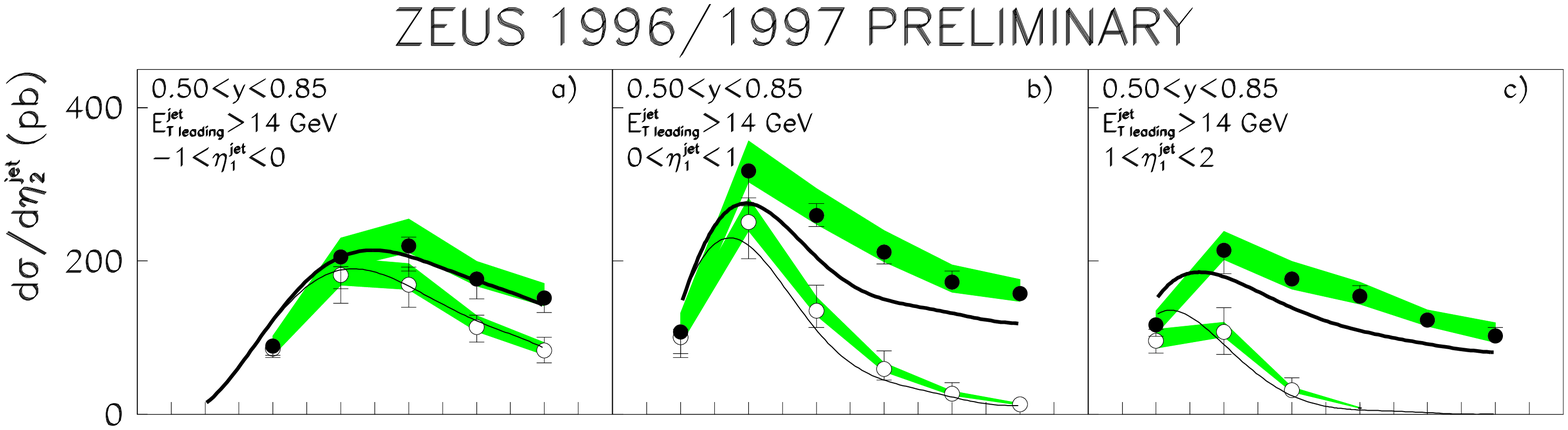}}
\put(0,120){\includegraphics[width=2.56in]{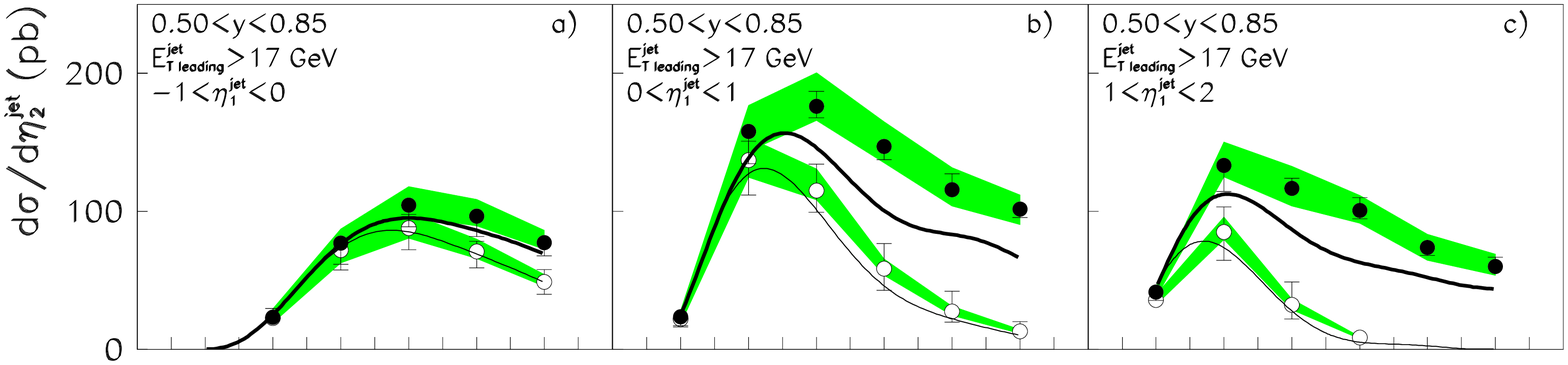}}
\put(0,080){\includegraphics[width=2.56in]{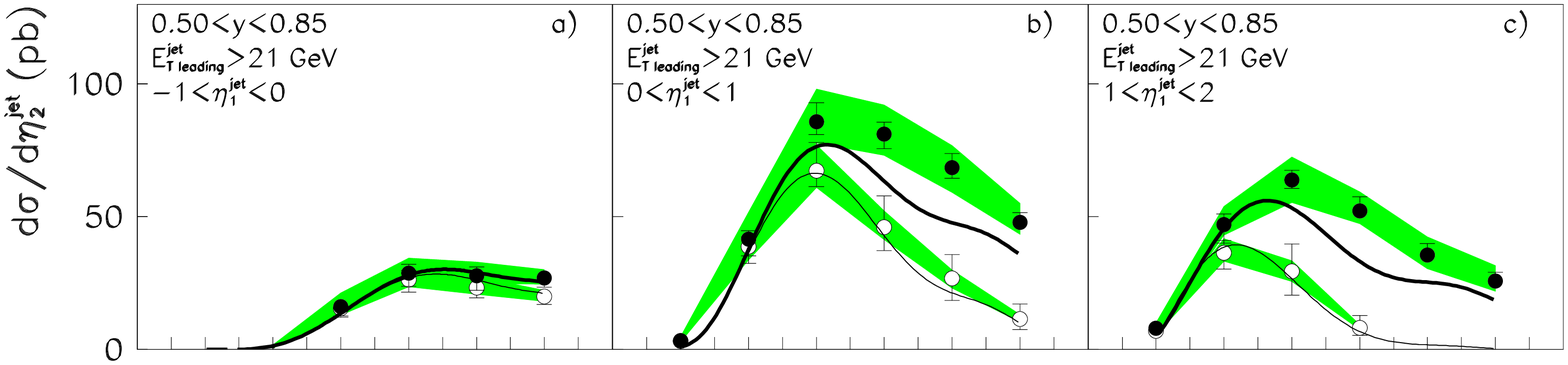}}
\put(0,040){\includegraphics[width=2.56in]{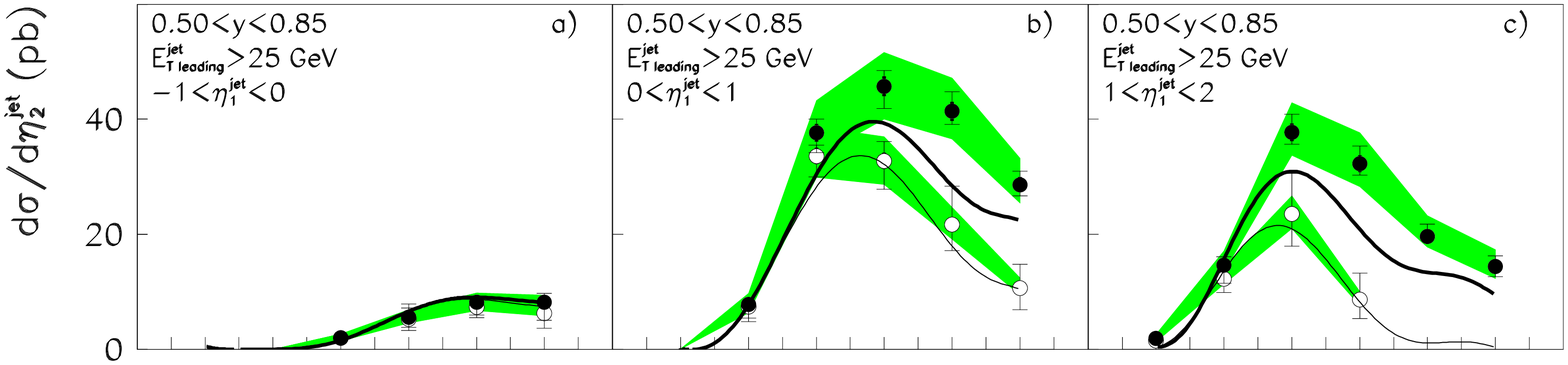}}
\put(0,000){\includegraphics[width=2.56in]{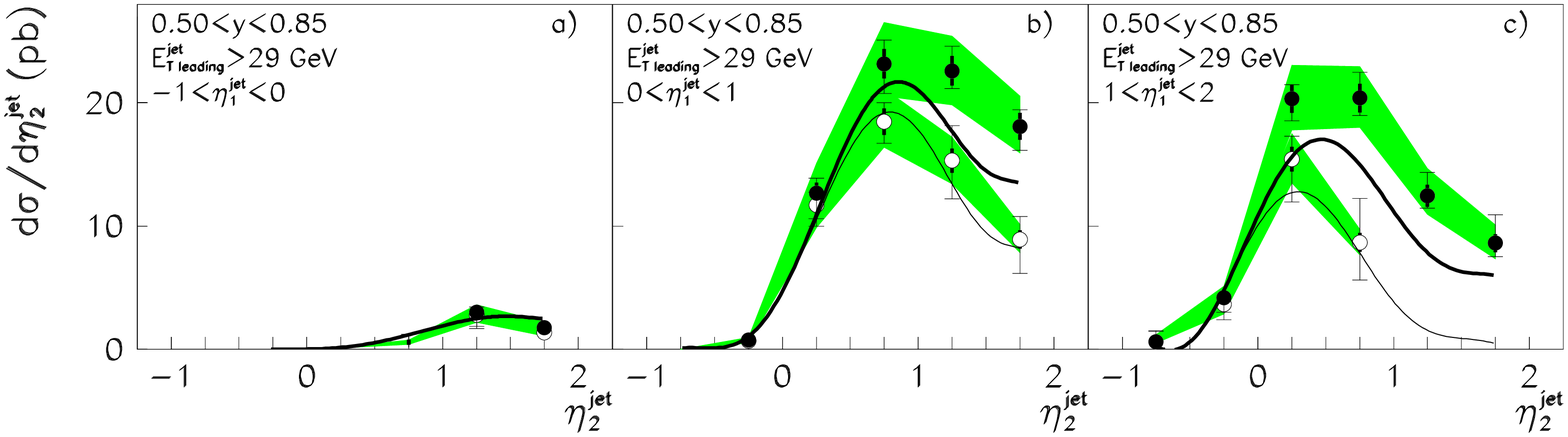}}
\put(19.5,-.5){\includegraphics[width=.6in]{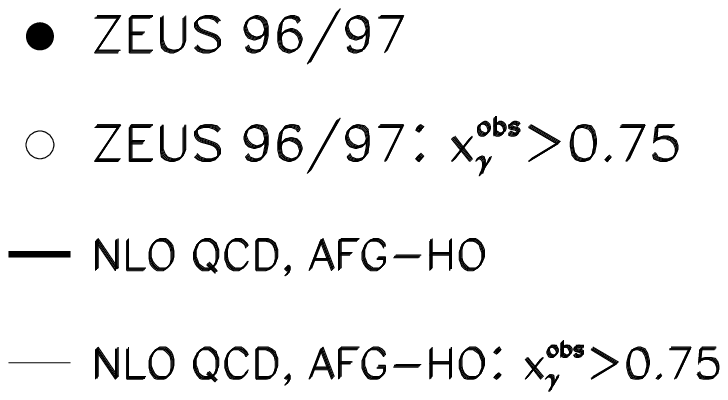}}
\end{picture}
\end{center}
\caption{\footnotesize Dijet cross sections as a function of the jet 
pseudorapidities, for $0.50<y<0.85$ and for different thresholds on the 
transverse energy of the leading jet, compared to NLO QCD.}
\label{fig:dsigdetanew_highy}
\vspace{-.5cm}
\end{figure}

The cross section as a function of the jet pseudorapidities 
(figure \ref{fig:dsigdetanew_highy}) was determined for thresholds 
on the highest transverse jet energy, ranging from 14~GeV to 29~GeV. 
To obtain a better correlation between the jet pseudorapidities 
and $x_\gamma$, the measurement was restricted to a kinematic region 
with: $0.50<y<0.85$. For the direct-enriched sample ($x_\gamma^{obs}>0.75)$ 
the data agree with NLO QCD. For the full $x_\gamma^{obs}$ range, 
however, at forward and central pseudorapidities, 
the cross sections lie above the calculations. This difference remains 
up to high transverse energies.  

Various theoretical uncertainties, that can affect the comparison between 
data and theory, were studied and found to be small 
(see also ~\cite{zeus_dij3}). 
Therefore the observed discrepancies, which are also present when 
other parametrisations of the photon structure are compared to, 
suggest that available parametrisations of the photon structure are 
too low in the region of high $x_\gamma$ and high factorisation scales.

\section{Summary}
Both HERA experiments, ZEUS and H1, have produced new results on 
the partonic structure of the photon. These data are complementary 
to $F_2^\gamma$ 
data from $e^+e^-$ experiments, and should 
be included in QCD fits used to determine parametrisations of the 
parton densities in the photon.

% A useful Journal macro
\def\Journal#1#2#3#4{{#1} {\bf #2}, #3 (#4)}

% Some useful journal names
\def\NCA{\em Nuovo Cimento}
\def\NIM{\em Nucl. Instrum. Methods}
\def\NIMA{{\em Nucl. Instrum. Methods} A}
\def\NPB{{\em Nucl. Phys.} B}
\def\PSNPB{{\em Proc. Suppl. Nucl. Phys.} B}
\def\PLB{{\em Phys. Lett.}  B}
\def\PRL{\em Phys. Rev. Lett.}
\def\PRD{{\em Phys. Rev.} D}
\def\ZPC{{\em Z. Phys.} C}
\def\EPJ{{\em Eur. Phys. J.} C}
\def\CPC{\em Comp. Phys. Commun.}

\end{document}